PREPRINT

# A Modular System Architecture for an Offshore Off-grid Platform for Climate-neutral Power-to-X Production in $H_2$Mare

Pascal Häbig[a]*, Daniel Dittler[b], Maximilian Fey[a], Timo Müller[b], Nikola Mößner[a], Nasser Jazdi[b], Michael Weyrich[b], Kai Hufendiek[a]

[a]*University of Stuttgart, Institute of Energy Economics and the Rational Use of Energy (IER), Heßbrühlstraße 49a, 70565 Stuttgart, Germany*
[b]*University of Stuttgart, Institute of Industrial Automation and Software Engineering (IAS), Pfaffenwaldring 47, 70550 Stuttgart, Germany*

* Corresponding author. Tel.: +49 (0)711 685 60901; *E-mail address:* pascal.haebig@ier.uni-stuttgart.de

**Abstract**

Power-to-X (PtX) products constitute a promising solution component in the defossilisation of hard-to-abate sectors. Where direct electrification is not possible, they can find application as chemical feedstock, easy to transport energy carriers and storage. In this contribution, a modular system architecture for a highly automated operation of an offshore off-grid PtX production platform consisting of five layers (asset, control, communication, digital, application) is presented. The assumed advantage of such an off-grid platform is the good scalability worldwide as well as easy exploitation of renewable potentials in any location. Using the example of a power-to-methanol process, a simplified prototype is outlined.

## 1. Introduction

In addition to the further expansion of electricity generation from renewable sources, the electrification of most energy applications is necessary to reach national and European climate targets. Where direct electrification is not possible, difficult to realise or just costly, e.g. in parts of the chemical industry or in certain segments of the transport sector, green hydrogen and other Power-to-X (PtX) products including methanol, ammonia and synthetic fuels attain importance, both as energy carriers of high energy density and as chemical feedstock. (e.g. [1-4])

A large number of research projects are currently investigating green PtX production and its further utilisation [5]. But only a few projects (e.g. FARWIND) focus on the offshore realisation of PtX processes [6]. Off-grid production systems powered by wind have the advantages of immediately utilising the wind power at the point of generation and eliminating the need for a capital-intensive grid connection. Also, transmission losses on the electricity side can be reduced.

This enables a wide spread realisation at spots with good renewable energy potentials worldwide without the need to realise long electric lines as sea transport of PtX products is flexible and cost efficient especially for long distances.

On the other hand, PtX processes and plant technologies are typically not designed for maritime environments and enabling offshore-capabilities incurs additional expenses. A further challenge is posed by the intermittency of wind power in case it is used as an exclusive energy source, introducing the need for energy storages, flexible to operate processes and advanced concepts for optimal process scheduling. The latter furthermore needs to consider the logistics of product offtake in case vessels are used instead of a pipeline. A high degree of automation is necessary to reduce the requirement for personnel on site. [6]

The purpose of this contribution is to present a modular system architecture for an offshore off-grid PtX production platform as a solution approach to the outlined challenges. The system components required to implement such a production platform are described generically and placed in the architecture. The overall objective is to promote the development of climate-neutral PtX production.

For context and background, different setups for PtX production systems and current research activities for offshore off-grid production systems are presented (Section 2). Subsequently, a modular system architecture for a PtX production platform is described and the six system components are associated to the five architecture layers (Section 3). Based on this, the implementation of a simplified prototype for a power-to-methanol process is presented. This is supposed to be used for further research (Section 4). The contribution closes with a summary and an outlook (Section 5).

## 2. Different Setups for PtX Production Systems Using Offshore Wind Power

Setups for PtX production systems using wind power as source for electricity can be differentiated based on the grid connection (on-grid vs. off-grid) and the supply of hydrogen (centralised vs. decentralised).

Today's production system setups consist of process equipment located onshore connected to the main grid infrastructure. An offshore wind farm will be simultaneously connected to an electrolysis plant, a PtX synthesis plant and to the grid infrastructure onshore via a suitable transmission infrastructure.

On-grid systems are the standard today and benefit from a steady power supply as well as the largest operational flexibility and product marketing options [7]. In periods of high electricity prices, power from the wind park could be fed into the grid. The wind park and electrolysis could also generate additional revenues through the provision of ancillary services, such as control reserve [8].

### *2.1. Different Production System Setups*

Power sourcing accounts for the most significant cost factor in PtX production. This leads to the consideration of saving the investment costs for the connection infrastructure (e.g. submarine cables and converter stations) resulting in an off-grid system, which is not connected to the onshore grid infrastructure [9].

In a fully offshore off-grid production system setup, both the electrolysis and the subsequent PtX synthesis steps are performed centrally on a single offshore production platform.

Leveraging the knowledge of platform construction from the oil and gas industry and taking advantage of the high flexibility of the electrolysis unit, it can also follow the power generation profile of the wind farm. The final PtX products are transported to the off-taker via pipeline or vessels. This kind of production system setup, which is explored in the German flagship project $H_2$Mare PtX-Wind [10], will be described in more detail in the further course of this contribution.

Offshore hydrogen production could also take place decentrally, with an electrolyser integrated into each wind turbine generator [11]. The hydrogen produced is collected on a central platform where PtX synthesis is conducted.

A decentral hydrogen supply is expected to reduce transmission losses and achieve higher reliability [9]. This is, for example, explored in the German $H_2$Mare OffgridWind hydrogen flagship project [10].

Off-grid systems offer a higher flexibility with regard to the choice of location [12]. This makes it possible to develop remote areas worldwide, such as offshore locations using floating systems [13]. Such sea areas are considered attractive due to much better wind conditions and less competition for land use. Furthermore, off-grid operation could eliminate the need for power transmission infrastructure and the associated investments. It can also improve the efficiency of the overall system by reducing transmission losses [9].

The described off-grid production system setup could also be transferred to non-marine environments with abundant renewable energy resources, such as remote locations on land with a high solar photovoltaics or wind power potential.

### *2.2. Research Activities for Offshore Production Systems*

The current focus of research activities for offshore production systems includes the improvement of individual subcomponents, e.g. by increasing the efficiency of various electrolysis technologies or by facilitating dynamic process operations, but also the development of offshore capabilities of process equipment, e.g. through the application of corrosion-resistant materials [5].

The German flagship project $H_2$Mare PtX-Wind explores the offshore off-grid production of green hydrogen and chemicals holistically. It combines research on processes and technologies with the development of operation concepts and sustainability assessments. For this purpose, concepts for three potential future production platforms at different development stages (prototype, pilot, industrial scale) are investigated. The *prototype platform* is envisioned to host containerised plant modules for PtX processes in an experimental scale on a floating pontoon in a maritime-like environment. The aim of the next bigger *pilot platform* is to build up the infrastructure to test the process components, plant modules, and entire process chains for PtX production under real offshore conditions. The insights from the prototype and pilot platforms can eventually, in combination with techno-economic analyses and life cycle considerations, lead to an economically viable concept for an industrial scale *production platform*. [14]

## 3. Modular System Architecture for an Offshore PtX Production Platform

The maritime environment and potentially remote location pose a number of challenges to the design and operation of offshore off-grid PtX production platforms. In order to address the challenges and requirements outlined in Section 1, a modular system architecture for a production platform was developed and will be presented below. Moreover, the generic system components are associated to their respective layer(s).

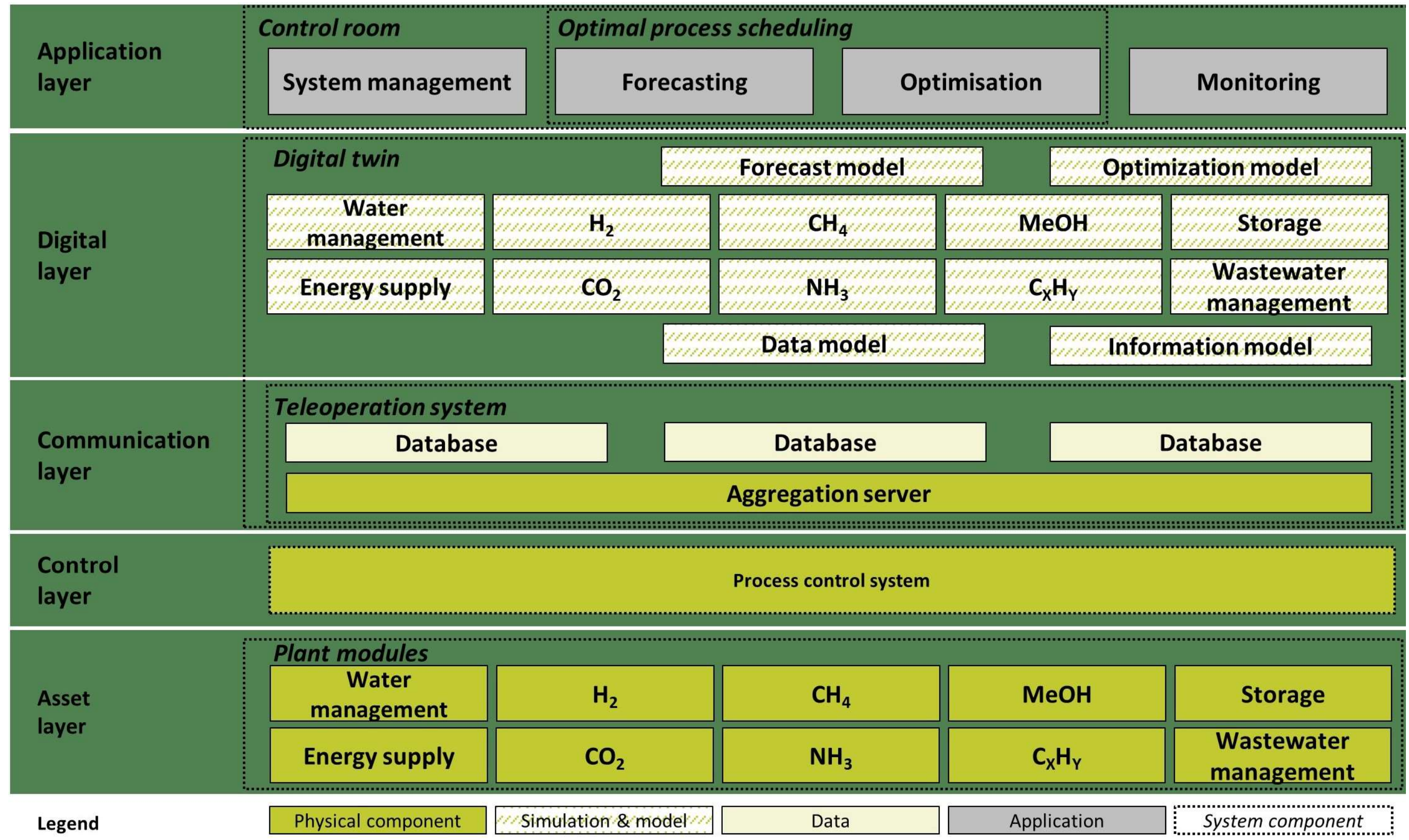

**Figure 1.** System architecture for a modular production platform with five layers based on [15].

### 3.1. Layered Modular System Architecture $H_2$MA

Following established architectures, e.g., the Reference Architecture Model Industrie 4.0 [15] or the Smart Grid Architecture Model [16], the $H_2$Mare architecture ($H_2$MA) consists of five layers: the asset layer, the control layer, the communication layer, the digital layer and the application layer. The individual layers are briefly explained below and illustrated in Figure 1.

The *asset layer* is the lowest layer and represents the foundation of the architecture. This layer contains all the physical plant modules that are necessary for the PtX production. The *control layer* records the process and environmental conditions with suitable sensors. It can also intervene in the process with appropriate actuators and thus control it. The *communication layer* contains the technologies and protocols required for data and information transmission. In the case of $H_2$MA, this is an aggregation server and a database for storage and filtering of data.

The *digital layer* is an image of the physical plant modules of the production platform. It is expected that the process modules will be obtained from different manufacturers and suppliers. Therefore, the digital image will contain a heterogeneous model landscape with data models, information models and behavioural models.

A *data model* is a concept used to organise and represent data in a structured way. It defines the relationships and properties of data to facilitate its use in applications and systems [17]. An *information model*, on the other hand, describes not only the structure of the data, but also its meaning and context. It is concerned with how information is processed within a system or organisation by providing a formal representation of the relevant concepts and relationships [18]. A *behavioural model* describes the system or subsystem in terms of its static or dynamic temporal behaviour in a quantitative form [19].

The individual applications, such as plant management, forecasting, optimisation, monitoring or virtual commissioning, are located on the *application layer*. For the execution of an application, recourse is made to subordinate layers, e.g. to models from the digital layer.

### 3.2. System Components of a Production Platform

The system architecture provides a generic framework for a production platform. In the following, the *system components* of such a platform are placed within it. To this end, the system components will be presented first and then related to the system architecture. The six *system components* include: plant modules, process control system, teleoperation system, digital twin, optimal process scheduling and control room.

*Plant modules,* e.g. water management, electrolysis ($H_2$), methanol-synthesis (MeOH), product storage, represent the technical equipment required for the PtX process chain. All the physical components of a production platform are located in the asset layer.

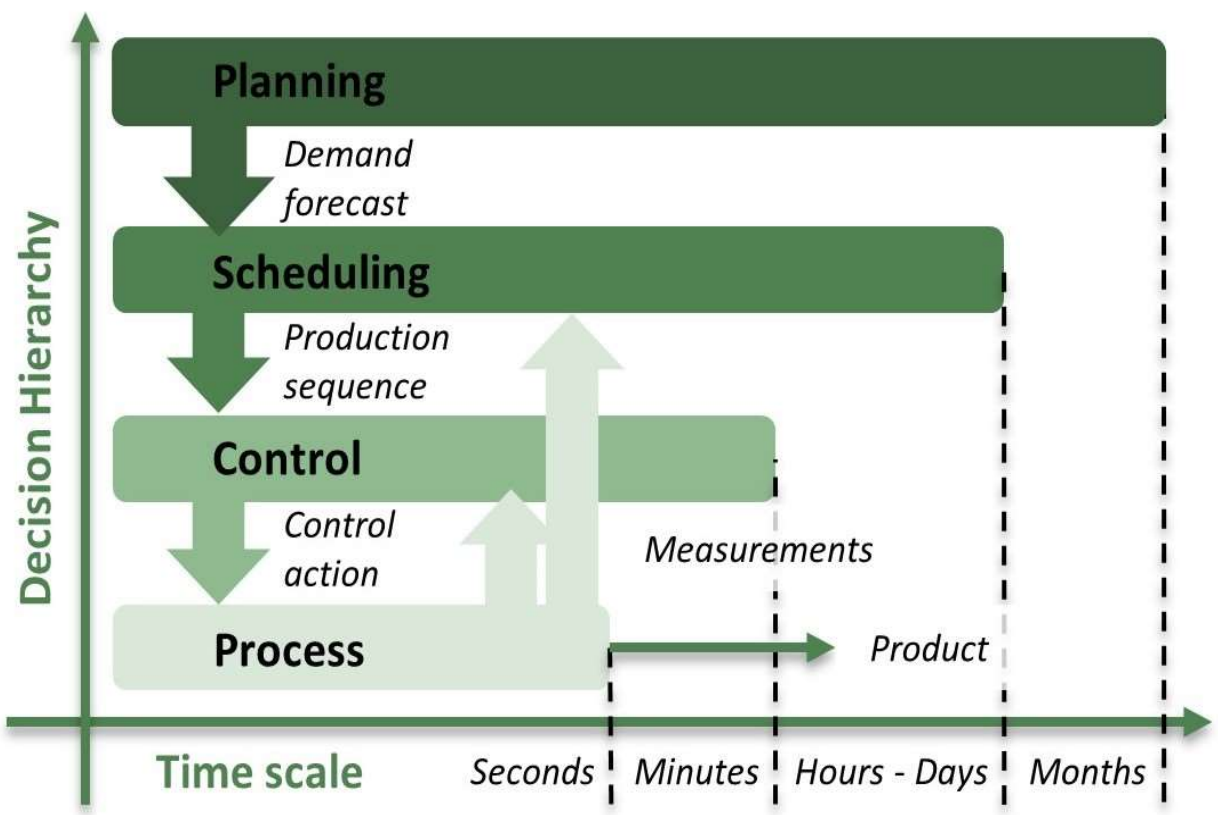

**Figure 2.** Decision hierarchy the operation of a production platform based on [24].

The *process control system*, that is equal to the control layer, integrates the individual plant modules into a higher-level control system. It enables an exchange of information across plant modules (e.g. storage level, temperature, energy and material flow).

The *teleoperation system* (communication layer) enables location-independent access to the process control system and the execution of applications such as remote monitoring or remote control.

In parallel to the physical plant modules of the production platform, a *digital twin* is a digital representation and serves as tool to perform various simulative experiments, such as simulations of different platform configurations and hardware-in-the-loop-based commissioning of plant modules.

In the literature, a significant increase of digital twin publications can be observed in the last years [20]. This has also given rise to a large number of different definitions in various fields [21].

In the authors' point of view, a digital twin, in the context given here, should be able to accommodate different interests (e.g. different simulative experiments) throughout the lifecycle of the production platform. For this contribution the definition of a digital twin that is summarised in [22] and based on [14] and [23] is utilised.

A digital twin is essentially composed of models and their interrelations. It advances from a mere digital replica by incorporating attributes described below.

Firstly, the models and their relationships are synchronised with the actual physical asset, which ensures an accurate representation of reality at any given time. Secondly, the digital twin must have a continuous flow of data acquisition from the physical asset, so that the dynamics can also be captured. Lastly, to accurately represent the dynamic behaviour of the physical asset, an executable model is necessary to complement the static image of the digital twin. [22]

Therefore, the digital twin in Figure 1 extends from the communication layer to the digital layer. The active data acquisition of the digital twin is realised by an aggregation server, which contains an information model. Operation data can be stored in connected databases with stored data models.

The various model types contained in the digital layer are used by different stakeholders over the lifecycle of a production platform. For example, individual process components (e.g. reactors), single plant modules (e.g. for a Fischer-Tropsch synthesis) or entire process chains (e.g. from electrolysis to methanation) are represented in individual behavioural models. Also these models are developed in different levels of detail and with different domain focus (e.g. fluidics or electrics). [19]

The system component *optimal process scheduling* is composed of the two applications forecasting and optimisation. It is located on the application layer. The overall objective is to maximise product output for a certain period of time or minimise production costs for a given output.

As shown in Figure 2, the decision hierarchy for the operation of a production platform can be described by a model of four levels: planning, scheduling, control, and process. Each level refers to a different time scale. On the highest level, a long-term production *plan* spanning over months is produced, which contains goals for production quantities and product quality. Based on the production plan, the *scheduling* intends to ensure an optimal platform operation for time scales between the next hours and the next few days. The *optimal process scheduling* is located on this level of the decision hierarchy. Operation set points resulting from the scheduling are communicated to the *control*. Process dynamics (*process*) in the time scale between seconds and minutes are handled on this level, for example, through the application of feedback loops. The *scheduling* receives feedback from the controlled system with updated system states as well. Nonlinearities and high dimensionalities are seen as major challenges that affect the computation time of the underlying process models. [24]

In a *control room*, operators monitor all the plant modules as well as the entire process chain, evaluate status messages and check alarm conditions. They are responsible for the overall system management and can always override operation set points suggested by the automated optimal process scheduling. These functions can be placed in the application layer of the architecture.

## 4. Development of a Simplified Prototype Using the Example of a Power-to-Methanol Process

The modular system architecture with its generic system components described in Section 3 serves as the foundation for further research activities. The objective is to develop a simplified prototype for an optimal process scheduling for an offshore off-grid production system.

For this purpose, the example of a power-to-methanol process will be modelled. A simplified prototype will allow for investigations, e.g. in the area of dynamic process optimisation. The analysis of probabilistic aspects related to the wind power forecast uncertainty will be of particular interest.

Ideally, the prototype of the architecture can eventually be applied and tested with a real process chain. In the development phase, the plant modules of the methanol synthesis process chain will be simulated digitally using the process simulation software *AVEVA Process Simulation*. The simulated process chain, as shown in Figure 3 consists of the plant modules seawater desalination ($H_2O$), electrolysis ($H_2$), methanol synthesis (MeOH) and product storage.

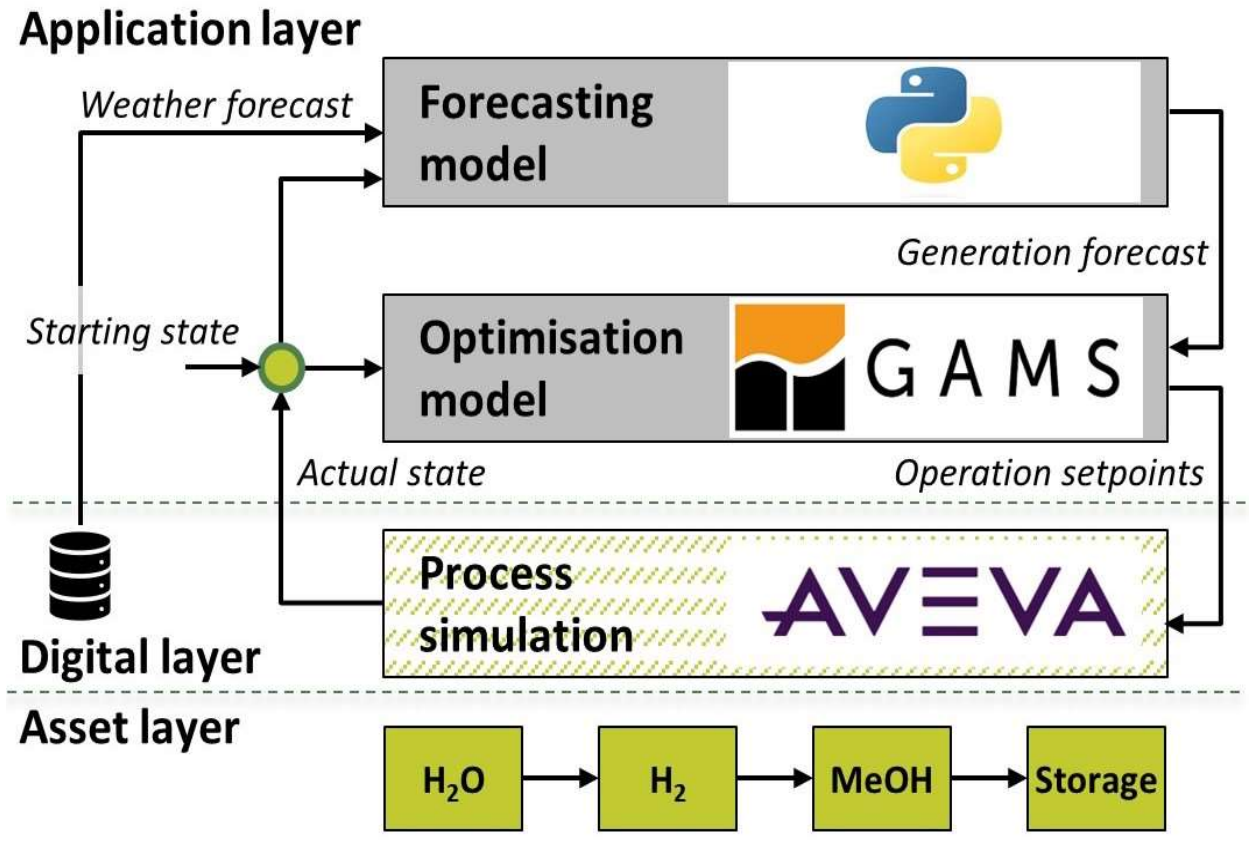

**Figure 3.** Simplified prototype for an optimal process scheduling using the example of a power-to-methanol process.

To perform optimal process scheduling, a wind power generation forecast will be computed automatically at periodic time intervals based on the weather forecast. The relationship between the relevant parameters of the weather forecast and the power generation will be provided by a machine learning model implemented in *Python*.

Besides the power generation forecast, the optimisation model will take the states of the plant modules (e.g. storage levels, mass and energy flows) as inputs. It will be implemented in GAMS, an algebraic modeling language.

With additional consideration of the logistics and maintenance plans, operating set points for various plant modules of the process chain will be calculated over a defined period of time and transferred to the process simulation interface.

Different scenarios, e.g. for low wind periods or periods with highly dynamic wind loads, will then be defined and simulation experiments will be performed with the simplified prototype.

## 5. Summary and Outlook

This contribution deals with the offshore off-grid platform for a climate-neutral PtX production in $H_2$Mare. PtX products constitute a promising solution component in the defossilisation of hard-to-abate sectors. Where direct electrification is not possible, they can find application as chemical feedstock, energy carriers and storage.

At the beginning, different setups for PtX production systems and current research activities for offshore off-grid production systems are presented. The production systems can be differentiated based on the grid connection (on-grid, off-grid) and the supply of hydrogen (centralised, decentralised).

The assumed advantage of such an offshore off-grid platform is the good scalability worldwide as well as easy exploitation of renewable potentials in any location.

To address the challenges that come with an offshore environment, a modular system architecture for a highly automated operation of an offshore off-grid PtX production platform is presented. The $H_2$Mare architecture ($H_2$MA) consists of five layers, namely asset layer, control layer, communication layer, digital layer and application layer.

Next, the six generic system components for a production platform are introduced. The functions of the plant modules, process control system, teleoperation system, digital twin, optimal process scheduling and control room are explained.

Based on this, the possible implementation of a simplified prototype for optimal process scheduling using a power-to-methanol process is outlined. This prototype is planned to be used for research on the following scientific questions.

With regard to the optimal process scheduling, particularly the interaction of wind power forecasts and operational optimisation, with a special focus on the impact of forecast uncertainties, bears potential for further research. In addition, a modelling approach to study the flexibility of the applied chemical processes or the entire off-grid production systems has to be investigated. In the context of the digital twin, the further research need is seen in model adaption to pick up the idea of "model-as-a-service" and to enable adaptive co-simulation. This requires, a suitable information model in order to be able to implement model coupling automatically.

In conclusion, a modular system architecture for PtX production systems can contribute to achieving commercial viability of such production systems and thereby lift large, so far untapped production potentials of PtX products, which are envisioned to play a central role in a future fossil free economy.

## Acknowledgement

This contribution was funded by the Federal Ministry of Education and Research (BMBF) under grant 03HY302R.